%% file: CALLPict.TEX
\documentclass [winedt,yap]{iitparc}
\usepackage{cite}
\usepackage{amsmath,amssymb,amsfonts,bm}
\usepackage{eufrak}
\usepackage{lscape}
\usepackage{floatfig,wrapfig,epsfig}
\usepackage{subfigure}
\usepackage{color}
\usepackage{rotating}
\usepackage{curves}
\usepackage{ulem}
\usepackage[mathscr]{eucal}
\usepackage{array}            
\RequirePackage{psfrag}
\RequirePackage{graphicx}

\begin{document}\normalem
\initfloatingfigs
\frontmatter          

\IssuePrice{25.00}%
\TransYearOfIssue{2017}%
\TransCopyrightYear{2017}%
\OrigYearOfIssue{2017}%
\OrigCopyrightYear{2017}%

\TransVolumeNo{78}%
\TransIssueNo{6}%
\OrigIssueNo{6}%
\OrigPages{157--172} 
\OrigCopyrightedAuthors{V.A. Malyshev, P.Yu. Chebotarev}


\mainmatter

\setcounter{page}{1087}
\CRubrika{INTELLECTUAL CONTROL SYSTEMS, DATA ANALYSIS} 
\Rubrika{INTELLECTUAL CONTROL SYSTEMS, DATA ANALYSIS}


%
\include{MalChebotarevTranslPict}

\end{document}

%% file: MalChebotarevTranslPict.tex

\def\xy{\hspace{.07em}}
\def\dE{\widetilde{d}_{\mathcal{E}}}                      
\def\dG{\widetilde{d}_{\mathcal{G}}}                      

%
%

\title{On Optimal Group Claims at Voting\\ 
       in a Stochastic Environment}
\author{V.A.~Malyshev$^{\xy*,**,a}$ and          
        P.Yu.~Chebotarev$^{\xy*,**,b}$          
        }
\institute{$^*$Trapeznikov Institute of Control Sciences, Russian Academy of Sciences, Moscow, Russia\\
           $^{**}$Moscow Institute of Physics and Technology, Moscow, Russia\\
           e-mail: $^a$vit312@gmail.com, $^b$pavel4e@gmail.com}

\received{Received October 28, 2016}
\authorrunning{MALYSHEV, CHEBOTAREV}
\titlerunning{ON OPTIMAL GROUP CLAIMS AT VOTING}

\maketitle

\begin{abstract}
There is a paradox in the model of social dynamics determined by voting in a stochastic environment (the
ViSE model) called ``pit of losses.''
It consists in the fact that a series of democratic decisions may systematically lead the society to the states unacceptable for all the voters.
The paper examines how this paradox can be neutralized by the presence in society of a group that votes for its benefit and can regulate
the threshold of its claims. We obtain and analyze analytical results characterizing the welfare of the whole society, the group, and the
other participants as functions of the said claims threshold.


\medskip\noindent$\!\!\!\!\!\!\!\!$\DOI{0091}
\end{abstract}

\setcounter{footnote}{1}

\section{Introduction}
\label{s_intro}

It is known that decisions made by the majority of voters on the proposals generated by a stochastic environment may systematically lead to situations unacceptable for all the voters. This voting paradox (the ``pit of losses'' paradox) has been described in~\cite{CheMal+16UBS}.
A natural approach to neutralizing this paradox is increasing the majority threshold (i.e., the number of votes necessary to accept a proposal).

However, there is another approach that deserves consideration.
Suppose that there is a ``group'' in the society whose participants pursue group rather than individual interests.
Moreover, the group has a ``claims threshold,'' i.e., the minimum profitability of proposals the group considers acceptable for it.
Which group claims threshold is profitable (optimal) for the whole society, for the group and for the other participants? In this paper, we give the answers to these and several other questions.

Consider the main features of the ViSE (Voting in Stochastic Environment) model~\cite{Che2706AiT}. Let \emph{society\/} consist of $n$ \emph{members\/}. $\ell$~members are \emph{egoists\/} and ${g = n - \ell}$ are \emph{group members}; therefore, ${\delta=\ell/n}$ is the proportion of egoists. The ViSE model extends the voting model introduced by A.V.\:Malishevsky (see~\cite{Mirkin74}), another extension of which is the well-known dynamic multidimensional voting model~\cite{McKelvey76}.

Each participant is characterized by the current value of his/her \emph{capital\/} (an alternative interpretation of it is \emph{utility}).
\emph{A proposal\/} (\emph{of the environment}) is a vector of proposed participants' capital increments. In the present paper, these increments are realizations of independent identically distributed normal random variables. The parameters of the corresponding normal distribution $N(\mu,\sigma^2)$ can be easily interpreted: the cases of ${\mu > 0}$, ${\mu = 0}$, and ${\mu <0}$ correspond to a favorable, neutral, and unfavorable environment, respectively; $\sigma$ characterizes the scatter of the proposals.

In the basic version of the ViSE model, the only stochastic agent is the environment. The behavior of the voters is deterministic in contrast to the random voting models~\cite{CoughlinNitzan81JET}.
An egoist votes for those and only those proposals that increase his/her capital. All members of a group vote for the proposals that are beneficial to this group and vote against the other proposals. The group can consider a proposal as a favorable one in the following cases: (a)~the average increase (which can be zero, positive, or negative) of a group member's capital exceeds the chosen \emph{claims threshold\/} $t$~\cite{Che09PU}, or (b)~the percentage of the group members receiving positive capital increments exceeds some threshold. In this paper, we consider the first case. The group supports those and only those proposals in which the average capital increment of its members is greater\footnote{This voting rule can be formulated in terms of the deterministic version of the model used in~\cite{PalfreyRosenthal85APSR}.} than a selected threshold~$t.$  Let $\alpha\in[0,1]$~be the strict relative \emph{majority threshold\/} for all votes of the society. If the proportion of the society supporting a proposal is greater than $\alpha,$ then the proposal is accepted and then implemented. The implemented proposals constitute the \emph{voting trajectory\/}. The subject of our study is the statistical dependence of such trajectories (which express the dynamics of social welfare) on the parameters of the model.

It should be noted that the behavior of the voters in the ViSE model corresponds to the Downsian concept~\cite{Downs57book}, more precisely, to its operationalization that is based on the comparison of the proposed state with \emph{status quo}~\cite{Grofman85JP}. Dynamic voting models have been studied in the theory of legislative bargaining~\cite{DugganKalandrakis12}, where in some cases
\cite{Penn09, DziudaLoeper14APSR, DziudaLoeper15} stochastic generation of proposals has been assumed. In this type of models, voters have ``ideal states'' that maximize their individual utility and the central problem is searching for equilibrium. Such kind of problems are trivial or even pointless in the ViSE model because the participants do not have finite ideal states. However, in this model, individual utilities (capitals) can be naturally aggregated into group utilities\footnote{Such aggregation is natural because of the assumption of the transferability of utilities.}.
This makes it possible to study collectivistic and altruistic voting strategies. Because of this, the ViSE model gives the researcher a tool
for studying cooperation and egoism, which provides an alternative to the simple games (such as Prisoner's Dilemma, Ultimatum, Avatamsaka, Public Goods, etc.) traditionally used for these purposes.

The ViSE model allows one to identify a number of social phenomena, mechanisms, and relationships associated with collective decisions. Their presence in practice (and degree) is a subject of special study. The model also allows to test various approaches to bringing society to the desired or
optimal state. Of course, this model can not reflect all aspects of social reality, however, all its predictions are characterized by mathematical transparency, therefore, the conditions for the manifestation of these phenomena are amenable to verification.

As indicated above, the paper examines the social dynamics in the presence of a group with a claims threshold.
A moderate group of this kind can be interpreted as an elite.

We obtain the expressions for the mathematical expectations of egoists' and group members' capital increments in the described society.
These expressions are presented in Section~\ref{s_capit}. In Section~\ref{s_coro}, it is shown that the group can choose such a claims threshold (it can be called \emph{optimal}) that the expected capital increment of the society is maximal. We provide a closed form expression for this threshold. The dependence of the participants' expected capital increments on the majority threshold is studied in Section~\ref{s_inc-th}. The influence of choosing the optimal claims threshold on the size of the ``pit of losses'' is examined in Section~\ref{s_opt-th-pit}.

\section{The capital increments of egoists and group members}
\label{s_capit}

    If the group does not support a proposal, then for its acceptance, the number of egoists' votes cast for it must exceed~$\alpha n.$
    Otherwise, if the group supports the proposal, then it is necessary and sufficient that the egoists give more then
    $$
      \alpha n - g
    = \alpha n - n + \ell
    = (\alpha + \delta - 1)n
    $$
    votes, i.e., the share (in the society) of the egoists supporting the proposal should exceed the value
\begin{gather}\label{e_gamma}
    \gamma = \alpha + \delta - 1.
    \end{gather}

    \begin{proposition}
    \label{pro:1}
    The mathematical expectation of egoist's capital increment $\dE$ in one voting step is equal to
    \begin{gather}\label{e_egoExp}
        \mathrm{M}(\dE)
       =\begin{bmatrix}\mu^+(\mu,\sigma,\ell,\gamma n)\quad \mu^+(\mu,\sigma,\ell,\alpha n)\end{bmatrix}
        \begin{bmatrix} P \\ Q \end{bmatrix}\!,
    \end{gather}
    where a row vector is multiplied by a column vector\/$,$ $\mu^+(\mu,\sigma,\ell,\ell_0)$ is the expectation of the normal voting sample of size $\ell$ with parameters $(\mu,\sigma)$ and voting threshold $\ell_0$ \emph{\cite{Che06AiT},}
    $P = F\left(\frac{(\mu - t)\sqrt{g}}{\sigma}\right),$ $Q = 1 - P,$ and $F(\cdot)$ is the standard normal distribution function.
    \end{proposition}

    The proofs of all results are given in the Appendix.

    \begin{proposition}
    \label{pro:2}
    The mathematical expectation of a group member's capital increment $\dG$ in one voting step is equal to
    \begin{gather}\label{e_groupExp}
        \mathrm{M}(\dG)
        =\begin{bmatrix}F_{\gamma n}\;\; F_{\alpha n}\end{bmatrix}
         \left(\mu \begin{bmatrix} P \\ Q \end{bmatrix} + \frac{\sigma f}{\sqrt{g}}\begin{bmatrix} 1 \\ -1 \end{bmatrix} \right)\!,
    \end{gather}
    where ${F_\xi = \sum_{x=[\xi]+1}^{\ell}\!b(x\,|\,\ell) \approx F\left(-\frac{[\xi]+0{.}5-p\ell}{\sqrt{pq\ell}}\right)},$
    ${b(x\,|\,\ell)=\binom{\ell}{x}p^xq^{\ell-x}}$ is the binomial probability function$,$
    ${p = F\left(\frac{\mu}{\sigma}\right)},$  ${q = 1 - p},$ 
    $P$ and $Q$ are defined in Proposition~$\ref{pro:1},$ ${f = f\!\left(\frac{(\mu - t)\sqrt{g}}{\sigma}\right)},$ and $f(\cdot)$ is the standard normal density.
    \end{proposition}

\begin{figure}[t]
\centering{\includegraphics[scale= 0.52,clip]{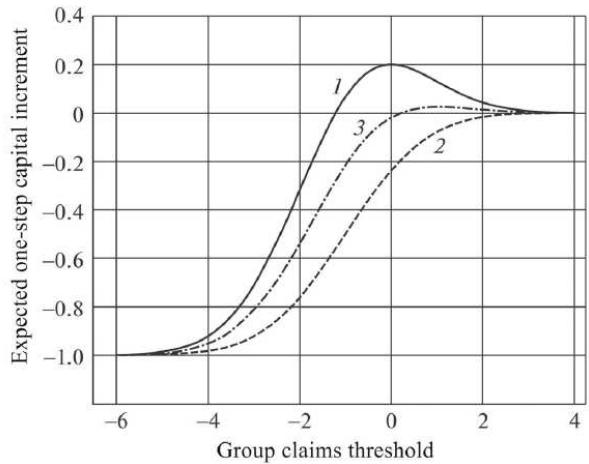}} 

\caption{\label{LargeGroup}
The expected one-step capital increments of a group member ({\sl1}\/), an egoist ({\sl2}\/), and a randomly selected participant ({\sl3}\/), where $n=100,\:$ $\ell=50,\: \mu=-1,\: \sigma=10,$ and $\alpha=0{.}5.$}
\end{figure}

Dependence of expected capital increments $\mathrm{M}(\dE)$ and $\mathrm{M}(\dG)$ of the participants on the claims threshold $t$ are shown in
Fig.\,\ref{LargeGroup}.

Fig.\,\ref{LargeGroup} demonstrates that in a moderately unfavorable environment ($\mu/\sigma=-0{.}1$), the group (half of the society in the present case) has a maximum income at zero claims threshold~$t.$ However, the egoists jointly lose more than the group wins, therefore, the expected capital increment of the whole society is negative, and so the society is gradually losing welfare. If the group's claims increase, then its income decreases more slowly than the loss of egoists. When the claims threshold $t$ is slightly higher than $0{.}2,$ the expected capital increment of the whole society becomes positive. When $t = 1$, the expected capital increment of the society reaches a maximum. A group with higher claims is blocking more and more profitable proposals, and the expected capital increment of the society is decreasing to zero.

\begin{figure}[t]
\centering{\includegraphics[scale=0.59,clip]{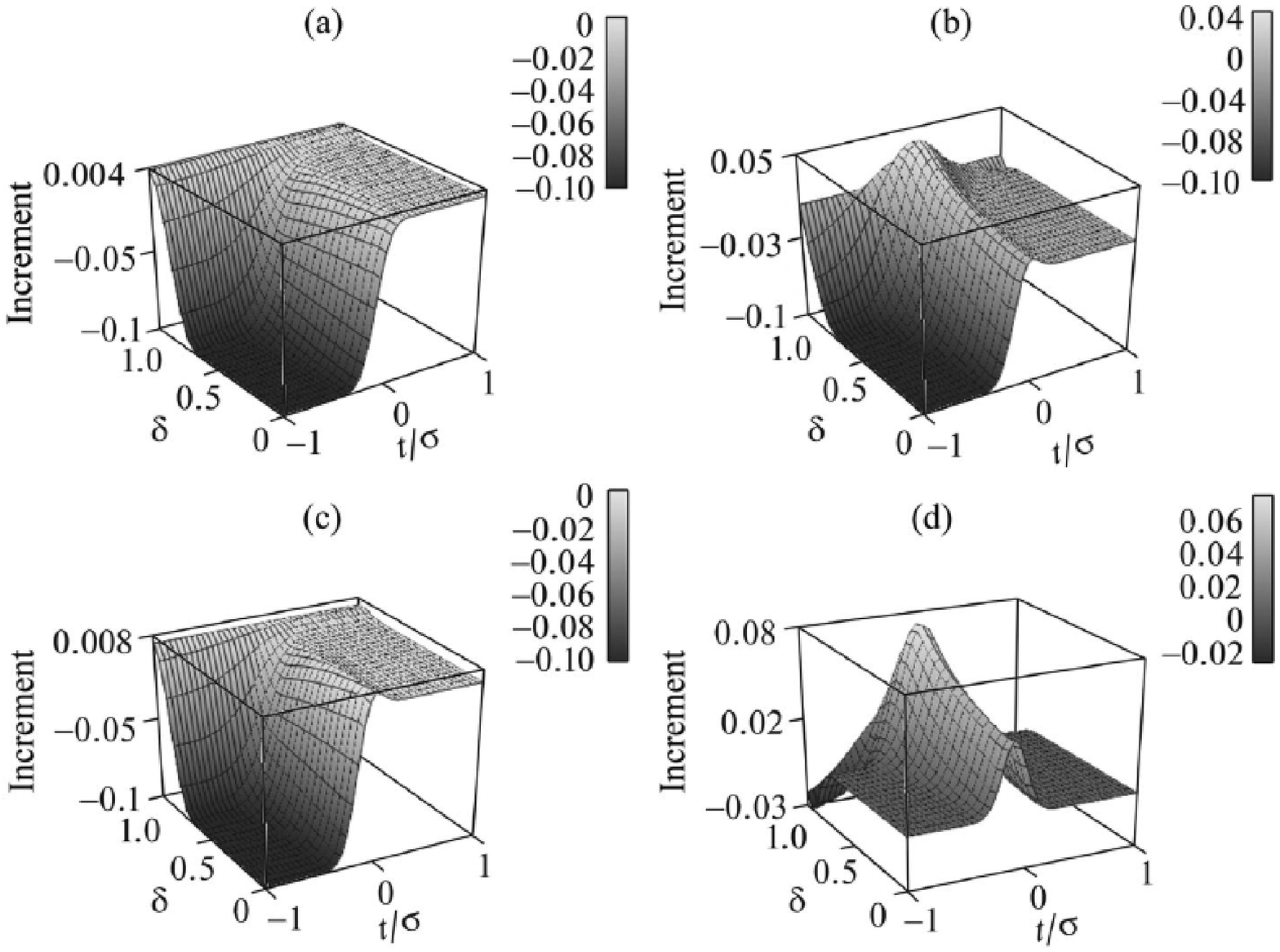}} 

\caption{\label{egoSurf} The expected one-step capital increments of an egoist~({\it a}\/), a group member~({\it b}\/), a randomly selected participant~({\it c}\/), and the difference between the expected capital increments of a group member and an egoist~({\it d}\/) as functions of the adjusted claims threshold of the group~$t/\sigma$ and the proportion of the egoists~$\delta,$ when $n=100$, $\mu=-0{.}1$, $ \sigma=1,$ and $\alpha=0{.}5.$}
\end{figure}

\begin{figure}[t]
\centering{\includegraphics[scale=0.59,clip]{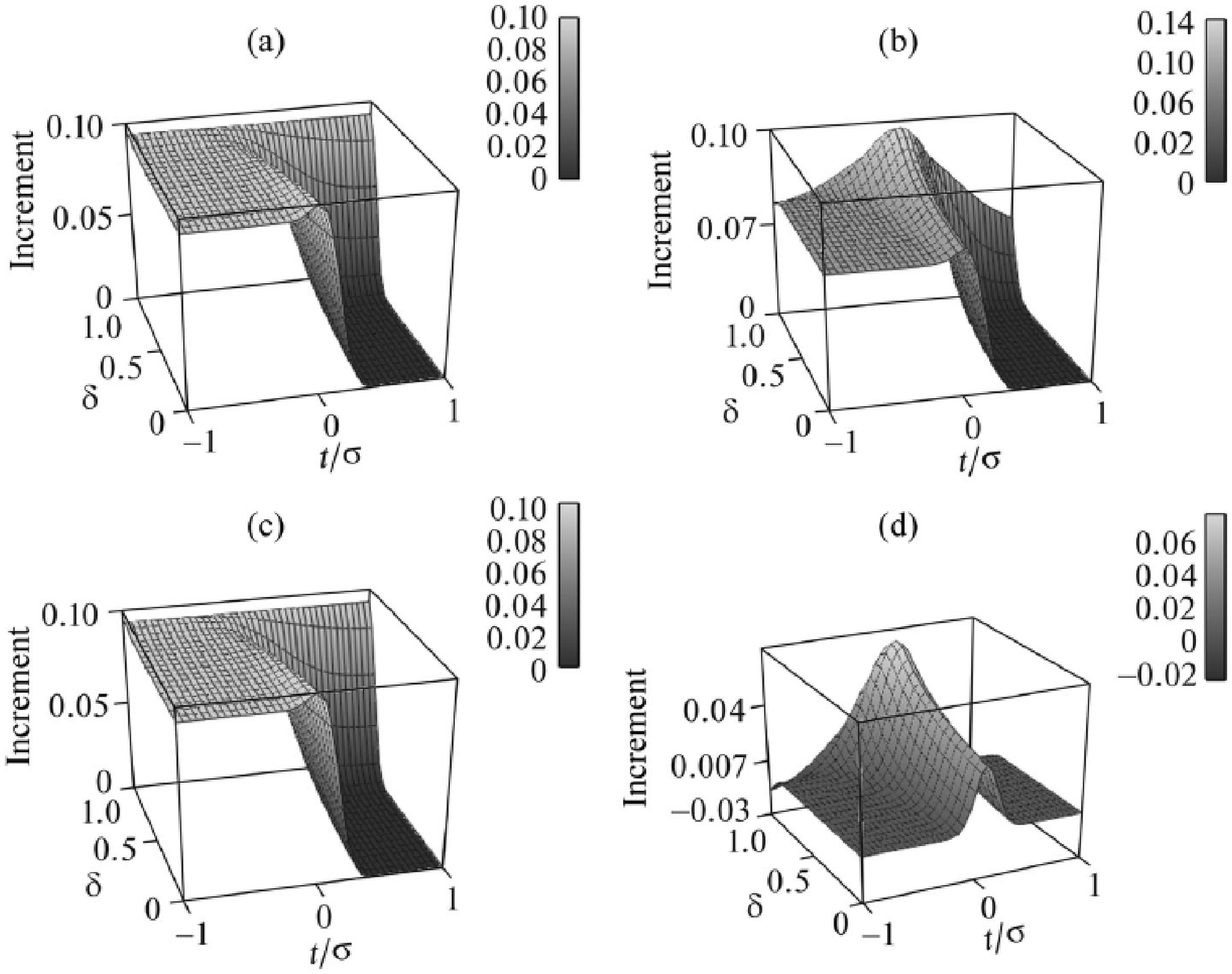}} 

\caption{\label{egoSurfPositive} The expected one-step capital increments of an egoist ({\it a}\/), a group member ({\it b}\/), a randomly selected participant ({\it c}\/), and the difference between the expected capital increments of a group member and an egoist ({\it d}\/) as functions of the adjusted claims threshold of the group $t/\sigma$ and the proportion of the egoists $\delta,$ when $n=100,\: \mu=0{.}1,\: \sigma=1,$ and $\alpha=0{.}5.$}
\end{figure}

\begin{figure}[t]
\centering{\includegraphics[scale=0.59,clip]{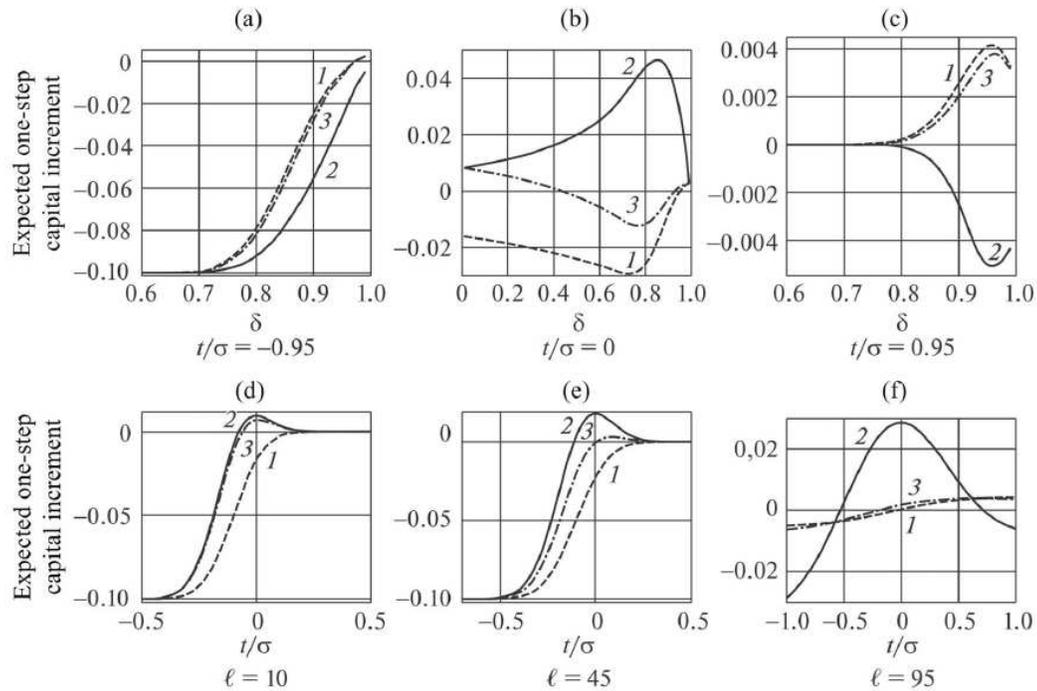}} 

\caption{\label{tSigma} The expected one-step capital increments of an egoist ({\sl1}\/), a group member ({\sl2}\/), and a randomly selected participant ({\sl3}\/): sections of the surfaces in Fig.\,\ref{egoSurf};\, $n=100,\: \mu=-0{.}1,\: \sigma=1,$ and $\alpha=0{.}5.$}
\end{figure}

Dependencies of the expected one-step capital increments of the participants on the adjusted claims threshold of the group
$t/\sigma$ and the proportion of the egoists $\delta$ are presented in Figures\:\ref{egoSurf} and \ref{egoSurfPositive}
for the cases of moderately unfavorable ($\mu/\sigma=-0{.}1$) and favorable ($\mu/\sigma=0{.}1$) environments. Fig.\,\ref{tSigma}
shows the sections of the surfaces presented in Fig.\,\ref{egoSurf}. It allows to compare the capital increments of an
egoist, a group member, and a randomly selected participant. These dependencies are characterized by the following regularities:
\begin{enumerate}
 \item[1.] For $\mu<0$ and any proportion of the egoists, the average capital increment of an egoist monotonically increases with the claims threshold of the group, because the number of accepted proposals that are unprofitable for egoists decreases. The fastest growth of the capital increment is observed at $t$ close to~$\mu$;
 \item[2.] When the proportion of egoists is small and the claims threshold of the group is low, the group accepts many unfavorable proposals.
This causes a fast loss of egoists' welfare;
 \item[3.] There is a maximum (with respect to the proportion $\delta$ of egoists) of the expected capital increment of an egoist in the domain of sufficiently high group's claims thresholds $t$ and a high proportion of egoists ($\delta>0{.}9$). For large $t,$ this maximum exceeds the average capital increment in the society consisting of egoists only.
This is caused by the fact that a small group with a high $t$ votes against most proposals. Therefore, the proposals must have a high support
from the egoists to be accepted. Such an actual increase of the majority threshold for egoists (for the approvement of a proposal, 51~votes are needed, but of 90--98 votes instead of 100) radically reduces the ``pit of losses'' \cite{CheMal+16UBS}, and the average capital increment of egoists
reaches the largest values;
 \item[4.] According to Theorem~1 in \cite{CheMal+16UBS}, for the society consisting of egoists and the environment with the parameters under consideration, the optimal majority threshold is $\alpha_0\approx0{.}52$;
 \item[5.] The group and the egoists lose welfare when the claims threshold and the proportion of egoists are low. When $t$ is close to $\mu,$ the expected capital increment of a group member increases rapidly with the growth of the claims threshold. However, in contrast to the average capital increments of the egoists, for any proportion of the group, the expected capital increment of a group member has a maximum at $t=0$ (since this threshold optimizes the group benefit). This maximum takes the greatest value when the group is small (cf.\:\cite{CLCLB09IPU});
 \item[6.] In the domain of high thresholds $t$ and the proportion of egoists $\delta>0{.}9,$ the expected capital increment of a group member has a minimum with respect to $\delta$, which is lower than the corresponding value for egoists. Thus, excessive claims of a small group worsen its position in comparison with the position of egoists;
 \item[7.] The expected capital increment of a group member is higher than that of an egoist everywhere, except for the domains of very high and very low claims thresholds and small group sizes.
\end{enumerate}

The expected capital increment of a randomly selected participant (surfaces ({\it c}\/) in Figures\:\ref{egoSurf} and~\ref{egoSurfPositive}) is expressed by the weighted average of the functions ({\it a}\/) and ({\it b}\/) with weights equal to the proportions of egoists and the group, respectively. The surface ({\it c}\/) has a combination of properties of the averaged surfaces.
A~randomly chosen participant loses capital when the group's claims threshold is low and the proportion of egoists is small. If the share of egoists is moderate, then the expected capital increment of an average participant (and of the whole society) has a maximum with respect to $t/\sigma$; the corresponding optimal group's claims threshold is found in Section~\ref{s_coro}. The value of the maximum is higher for a larger group. In comparison to the maximum for the group, this maximum is shifted to the domain of larger $t/\sigma,$ because if $t/\sigma$ grows from the group's optimum, then the expected capital increment of a group member decreases slower than the expected capital increment of an egoist increases. This leads to the growth of the expected capital increment of the whole society.

The established relationships allow us to draw the following conclusions.

The zero claims threshold is optimal for the group in any environment. It is advantageous for the group to be relatively small (about $15$ members out of 100 participants) in this case. In a moderately unfavorable environment, such a group becomes richer, whereas egoists lose welfare.

In an unfavorable environment, the optimal group's claims threshold is positive. The decrease of this threshold from the optimum causes a faster loss for the society then the increase from the optimum.

In a favorable environment, on the contrary, the optimal group's claims threshold is negative, and the expected capital increment of the society decreases faster when the threshold is deviated from the optimal value to the positive direction. This conclusion is analogous to the one of \cite{CheMal+16UBS} saying that the more favorable environment is encountered, the lower majority threshold is optimal.

\section{Optimal group's claims threshold}
\label{s_coro}

In this section, we obtain an analytical expression for the \emph{optimal threshold\/}, i.e., for the group's claims threshold that maximizes the capital increment of the society. Let $\beta$ be the ratio of the number of egoists to the number of group members:
$$
\beta=\frac{\ell}{n-\ell}=\frac{\delta}{1-\delta}.
$$
\begin{theorem}
\label{pro:opt} The expected one-step capital increment of the society reaches its maximum at the group's claims threshold
\begin{gather}\label{e_generalOptimT}
    t_0 = \frac{\beta}{F_{\gamma n} - F_{\alpha n}}\big(\mu^+(\mu,\sigma,\ell,\alpha n) - \mu^+(\mu,\sigma,\ell,\gamma n)\big),
\end{gather}
where the notations is introduced in Section~$\ref{s_intro}$ and Propositions~$\ref{pro:1}$ and~$\/\ref{pro:2}.$
\end{theorem}

Let us find out how the optimal threshold $t_0$ 
depends on the model parameters. Since $\gamma < \alpha$ (see\,\eqref{e_gamma}), it follows from the definition of $F_\xi$ that $F_{\gamma n} - F_{\alpha n} > 0$ (and the smaller the share of the group $1-\delta$, the smaller this difference); $\beta$~is nonnegative. Consequently, the sign of $t_0$ coincides with the sign of the difference $\mu^+(\mu,\sigma,\ell,\alpha n) - \mu^+(\mu,\sigma,\ell,\gamma n).$

\begin{figure}[b]
\centering{\includegraphics[scale=0.52,clip]{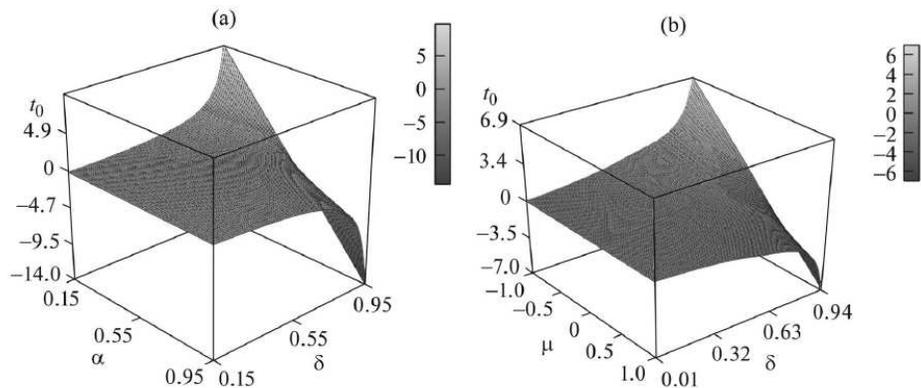}}

\caption{\label{t0Surf1} Dependence of the optimal claims threshold $t_0$ of the group on:
     ({\it a}\/) the proportion of egoists $\delta$ and the majority threshold $\alpha$ for $\mu=0{.}1,\: n=100,$ and $\sigma=1$;
     ({\it b}\/) the proportion of egoists $\delta$ and the mathematical expectation of the proposals $\mu$ for $\alpha=0{.}5,\: n=100,$ and $\sigma=1.$}
\end{figure}

Fig.\,\ref{t0Surf1},{\it a}\/ demonstrates the dependence of $t_0$ on the proportion of egoists~$\delta$ and the majority threshold~$\alpha$ for $n=100$ and $\mu/\sigma=0{.}1.$

We note that for a large group, the absolute value of threshold $t_0$ is small because the factors $\beta$ and $(F_{\gamma n} - F_{\alpha n})^{-1}$ are close to zero. The decrease of $t_0$ with the increase of the majority threshold~$\alpha$ becomes faster with the growth of the number of egoists. The shape of this surface does not substantially depend on the sign of~$\mu$.

Fig.\,\ref{t0Surf1},{\it b}\/ shows the dependence of $t_0$ on the proportion of egoists~$\delta$ and the mean of the environment proposals $\mu$ for $n=100,\: \alpha=0{.}5,$ and $\sigma=1.$

The two surfaces in Fig.\,\ref{t0Surf1} have the same shape. When~$\mu$ increases, $t_0$ decreases and the smaller the group is, the faster the change is.

\begin{figure}[t]
\centering{\includegraphics[scale=0.52,clip]{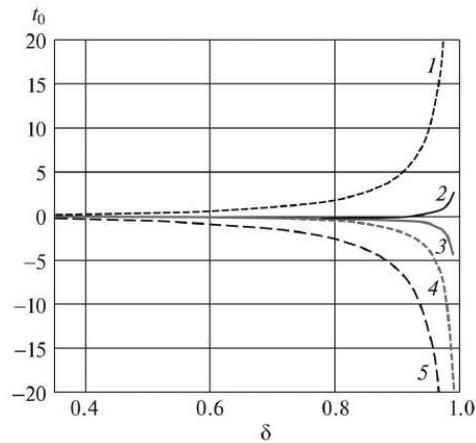}}

\caption{\label{t0} Dependence of group's optimal claims threshold $t_0$ on the proportion of egoists $\delta$ for various majority thresholds
($\alpha=0{.}15$~({\sl1}\/), $\alpha=0{.}46$~({\sl2}\/), $\alpha=0{.}5$~({\sl3}\/), $\alpha=0{.}6$~({\sl4}\/), and $\alpha=0{.}9$~({\sl5}\/)); $n=100,\:
\mu=0{.}1,$ $\sigma=1.$}
\end{figure}

Fig.\,\ref{t0} shows various sections of the surface presented in Fig.\,\ref{t0Surf1},{\it a\/} by the planes of fixed majority thresholds. For
${\alpha=0{.}46},$ we have ${t_0\approx0}$; for the smaller~$\alpha$'s, the optimal claims threshold~$t_0$ is positive and increases with~$\delta$; for the lager~$\alpha$'s, it is negative and decreases.

\begin{corollary*} 
\label{co:2} If the votes of a group are sufficient to accept a proposal $(\alpha<1-\delta),$ then
$$t_0=\frac{\beta}{1-F_{\alpha n}}(\mu^+(\mu,\sigma,\ell,\alpha n)-\mu).$$ 

If the votes of the egoists are insufficient to accept a proposal $(\delta\leqslant \alpha),$ then
$$t_0=-\frac{\beta}{F_{\gamma n}}\xy\mu^+(\mu,\sigma,\ell,\gamma n).$$ %

If both of the above conditions are met $(\delta\leqslant \alpha<1-\delta),$ then
$$t_0=-\beta\xy\mu.$$
\end{corollary*}

In the latter case, the optimal group's claims threshold $t_0$ has an extremely simple expression. If the number of votes necessary for a proposal to be accepted exceeds the number of egoists and does not exceed the size of the group, then $t_0$ is in the same ratio to $(-\mu)$ as the number of egoists to the number of group members is, and the larger the group is, the closer to zero $t_0$ is.

\section{Dependence of capital increments on the majority threshold}
\label{s_inc-th}

The expression of the optimal majority threshold for a society consisting of egoists has been obtained in~\cite{CheMal+16UBS}. Now we consider the problem of optimizing the majority threshold $\alpha$ for the society that consists of egoists and a group.
Recall that the optimal threshold is the threshold that maximizes the increment of the total capital of the society. 

Using Propositions~\ref{pro:1} and~\ref{pro:2} consider the analytical dependencies of the expected capital increments of an egoist, a group member, and a randomly selected participant on the majority threshold $\alpha$ and the proportion of egoists $\delta$ in the moderately unfavorable environment ($\mu/\sigma=-0{.}1$) for the zero group's claims threshold~$t.$
%
%

\begin{figure}[b]
\centering{\includegraphics[scale=0.52,clip]{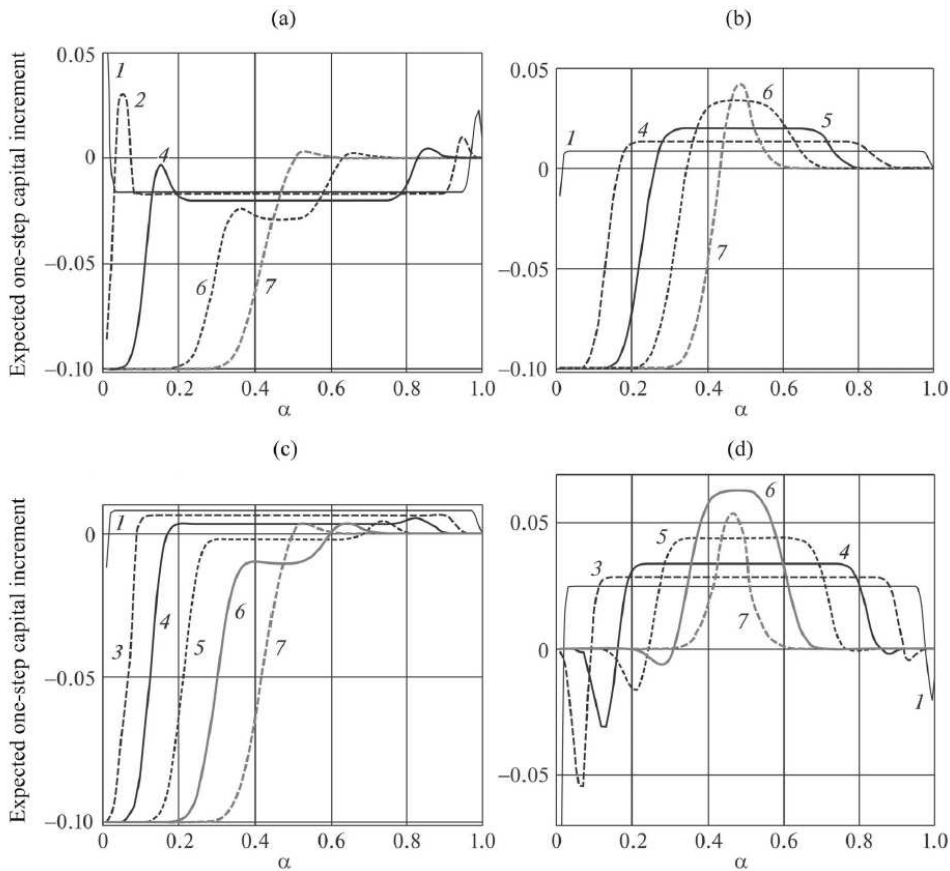}}

\caption{\label{deltaAlphaSurf} The expected one-step capital increments of an egoist ({\it a}\/), a group member ({\it b}\/), a randomly selected participant ({\it c}\/), and the difference between the expected capital increments of a group member and an egoist ({\it d}\/) as functions of the majority threshold $\alpha$ for $n=100,\: \mu=-0{.}1,\: \sigma=1,\: t=0,$ and various proportions of egoists in the society ($\delta=0{.}02$ ({\sl1}\/), $\delta=0{.}1$ ({\sl2}\/), $\delta=0{.}15$ ({\sl3}\/), $\delta=0{.}3$ ({\sl4}\/),
$\delta=0{.}5$ ({\sl5}\/), $\delta=0{.}7$ ({\sl6}\/), and $\delta=0{.}95$ ({\sl7}\/)).}
\end{figure}

\sloppy
The sections of the corresponding surfaces by the planes of equal egoists shares are shown in Fig.\,\ref{deltaAlphaSurf}. If~$\ell$ is not very high, then sections~({\it a}\/) have two maxima with respect to~$\alpha$; the curves of maxima arguments intersect at the point (${\delta=1}$, ${\alpha=\alpha_0}$), where $\alpha_0$ is the optimal majority threshold in the society consisting of egoists. Consider the reasons for the appearance of these ``ridges.''

\sloppy
We start with the relationship for egoists (Fig.\,\ref{deltaAlphaSurf},{\it a}). At a very low proportion of egoists and a low majority threshold, an egoist has a high expected capital increment, because the probability of the event that all the egoists are satisfied with a proposal is markedly different from zero, while their votes are sufficient to accept such a proposal.
Therefore, their expected one-step capital increment is comparable in the absolute value to~$\mu$. When the majority threshold is higher than the proportion of egoists, then they are unable to accept a proposal by their votes only, and their expected capital increment is much lower. The second maximum is realized at the majority threshold close to~$1$. Its appearance is caused by the following fact.
When the adoption of a proposal requires both the votes of the group and the egoists (i.e., overall, the egoists have a veto), then egoists' influence on decision-making is quite noticeable, and their interests are substantially taken into account. However, the support of the proposal by the group is mandatory for the realization of this scenario, while with $\mu<0,$ it does not happen very often, so this maximum is much lower than the 
first one.

If the number of egoists increases, then both maxima still exist, but they shift to $\alpha_0$ (with a linear dependence of their arguments on~$\delta$) and become lower. It can be simply explained. Here, one egoist capable of accepting a proposal is replaced by a ``clique'' which less often provides the required number of votes. The expected capital increment of a ``clique'' member (when the clique is satisfied with the proposal) is lower than that of the above egoist due to the law of large numbers.
The part of the curve behind the second maximum lies somewhat lower, since an excessive majority threshold does not even allow to accept proposals that are beneficial to both the group and most egoists.

The dependence for the group (Fig.\,\ref{deltaAlphaSurf},{\it b}) has two ``slopes'' on the $\delta$-sections. Between them, there is a ``plateau.''
It shows that the majority threshold located between the maxima of the expected capital increment of the egoists is quite beneficial for the group. Lower thresholds allow the egoists to ignore the interest of the group, which reduces the group welfare. Excessive thresholds do not even allow to accept highly profitable proposals.
As can be expected, the ``plateau'' is higher when the proportion of egoists is larger, because a smaller group has a greater benefit per participant.

Fig.\,\ref{deltaAlphaSurf},{\it c} depicts the expected capital increment of a randomly selected participant. As in the case of the egoists, the corresponding surface in the coordinates $\alpha,\,\delta$ has two ``ridges.'' When the proportion of egoists $\delta$ is high, the surface of the expected capital increment of a randomly selected participant is close to that of an egoist. The height of the maxima decrease as the number of egoists increases. At a low $\delta$, the surface of a randomly selected participant is close to the group surface and does not have high peaks.

Fig.\,\ref{deltaAlphaSurf},{\it d} shows the difference between the expected capital increments of a group member and an egoist. This difference is positive except for the domain near the first ``ridge'' of the egoists' surface, where they win, because the egoists can ``dictate their terms'' in this case.

\section{Claims threshold adjustment to reduce the ``pit of losses''}
\label{s_opt-th-pit}

\begin{figure}[t]
\centering{\includegraphics[scale=0.52,clip]{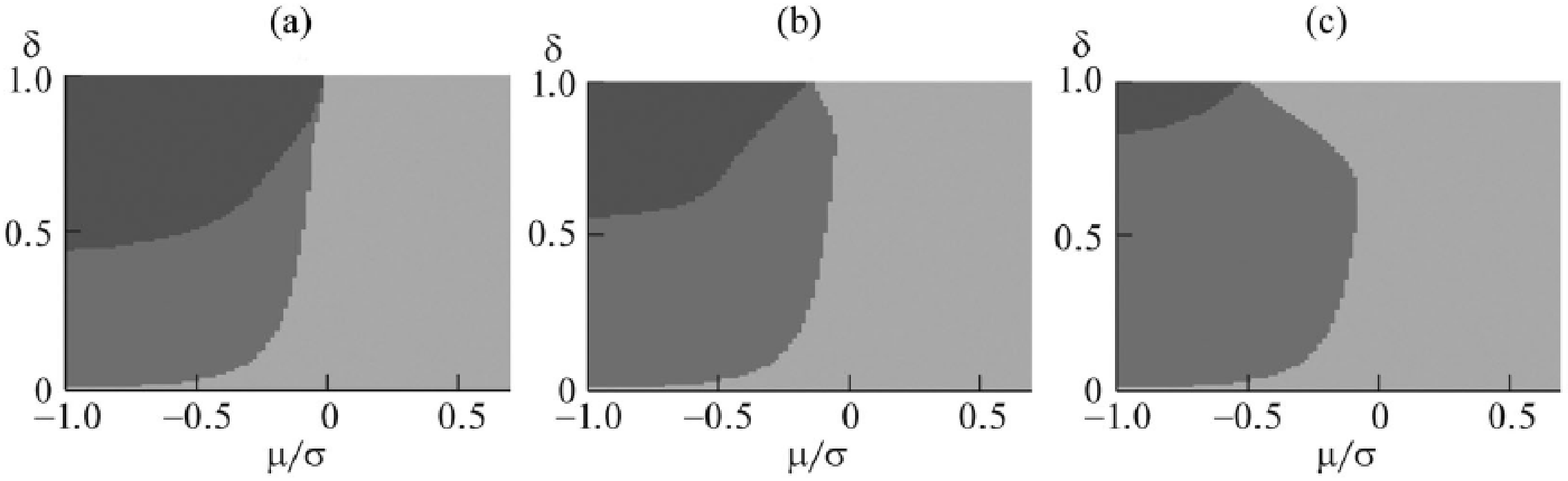}}

\caption{\label{pit} The influence of adopting the optimal claims threshold on the presence of the ``pit of losses'' for ${n=100}$.
    ({\it a}\/)~${\alpha = 0{.}4}$; ({\it b}\/)~${\alpha = 0{.}5}$; ({\it c}\/)~${\alpha = 0{.}6}$.
	The medium gray area is the area of the presence of ``pit of losses'' for ${t=0}$, the dark gray area corresponds to the optimal claims threshold ${t=t_0}$.
}

\bigskip
\centering{\includegraphics[scale=0.52,clip]{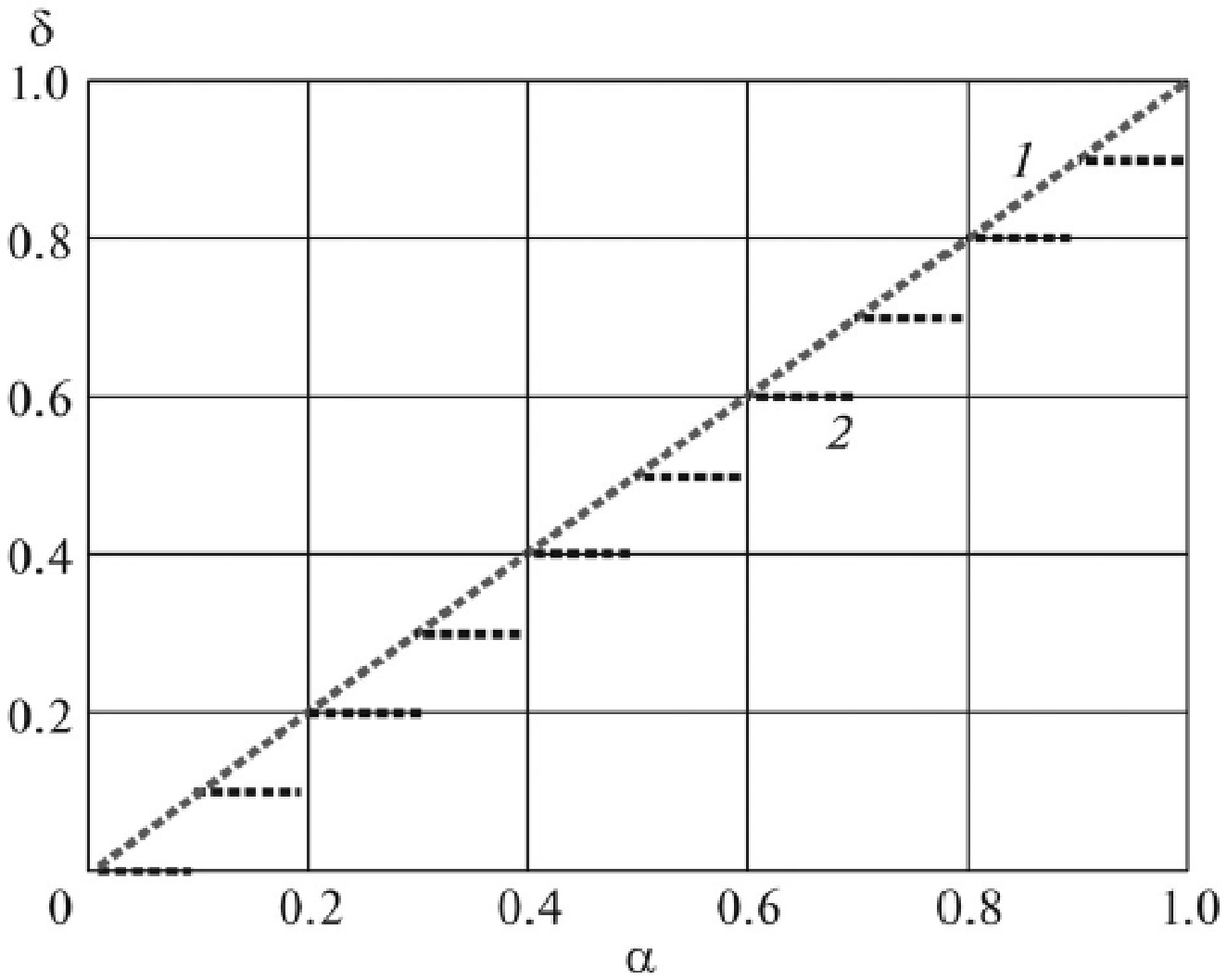}}

\caption{\label{maxDeltaFromAlpha} Dependence of the maximum proportion of egoists for which the choice of group's optimal claims threshold neutralizes the ``pit of losses'' on the majority threshold~$\alpha.$ Gray points correspond to $n = 100$ ({\sl1}\/), black points to $n = 10$ ({\sl2}\/).}
\end{figure}

The group can reduce the size of the ``pit of losses'' (the area with negative expected capital increments of the society) by choosing an optimal claims threshold. If the group has a sufficient size, then it can completely remove the pit. Fig.\,\ref{pit} illustrates this possibility. In other words, a sufficiently large group can help the society to overcome the paradoxical capital reduction caused by democratic decision-making. It is important to note that a higher majority threshold allows a smaller group to do so. For example, if $\mu/\sigma>-1,$ then for $\alpha = 0{.}4$ the ``pit of losses'' does not appear for the societies with the proportion of egoists up to $44\,\%$; for $\alpha=0{.}5,$ up to $56\,\%$, and for $\alpha = 0{.}6,$ up to $83\,\%$.

Fig.\,\ref{maxDeltaFromAlpha} shows the dependence on $\alpha$ of the maximum share of egoists for which the choice of the optimal group's claims threshold $t=t_0$ removes the ``pit of losses.'' This relationship is basically expressed by the function $y = x$ (after the correction taking into account the equivalence of majority thresholds that determine the same winning coalitions).

Thus, it is established that the choice of the optimal group's claims threshold can be considered as a tool to neutralize the ``pit of losses'' paradox.

\section{Conclusion}

In this paper, we considered the dependence of the group's claims threshold that maximizes the capital increment of the society on the parameters of the ViSE model for a society consisting of a group and egoists.
The absolute value of this threshold (called optimal) increases as the proportion of egoists in the society grows. If the percentage of egoists is fixed, then the optimal group's claims threshold decreases with the grow of the majority threshold and becomes negative.
Moreover, the optimal group's claims threshold decreases as the mean~$\mu$ of the proposals grows. The latter relationship is essentially a counterpart of one of the conclusions of~\cite{CheMal+16UBS}, which recommends increasing the majority threshold in deteriorating environments and reducing it in improving environments.

We have shown that the choice of the optimum group's claims threshold helps to neutralize the ``pit of losses'' paradox in the societies consisting of a group and egoists. For a complete success, the proportion of egoists should not exceed a certain value depending on the majority threshold.
Thereby, an increase of the majority threshold expands the zone of neutralization (by means of choosing the optimal group's claims threshold) of the ``pit of losses'' paradox.

\section*{Acknowledgments}
This work was supported by the Russian Science Foundation, project 16-11-00063 granted to IRE~RAS.

\appendix{}

\PPR{\ref{pro:1}}
Let $G$ be the event that the group supports a proposal, $\overline{G}$ is the opposite event, $P=P(G),\,$ and $Q=P(\overline{G}) = 1 - P.$

When $G$ occurs, it is necessary and sufficient for the proposal acceptance that egoists add more than $\gamma n$ votes (case~(a)). If the group
votes against the proposal, then more than $\alpha n$ votes of egoists are required to accept the proposal (case~(b)). Since the average of the independent capital increments of the group members is distributed as $N\big(\mu, \frac{\sigma^2}{g}\big),$ the probabilities of events $G$ and $\overline{G}$ are given by
\begin{gather}\label{e_PQ}
P=F\left(\frac{(\mu - t)\sqrt{g}}{\sigma}\right)\;\;\; \mbox{and}\;\;\;
Q=1 - P = F\left(\frac{(t -\mu)\sqrt{g}}{\sigma}\right).
\end{gather}

Formulas for the expected capital increments of an egoist in cases (a) and (b) can be obtained using Lemma on ``normal voting samples'' \cite{Che06AiT} and \eqref{e_PQ}, which leads to the expression~\eqref{e_egoExp}:
{\arraycolsep=.4mm\begin{eqnarray*}
        \mathrm{M}(\dE)
       &=&\mathrm{M}\big(\dE\,|\,G\big)P(G) + \mathrm{M}\big(\dE\,|\,\overline{G}\big)P(\overline{G}) 
       \\\nonumber
       &=& \mu^+(\mu,\sigma,\ell,\gamma n)\,F\!\left(\frac{(\mu - t)\sqrt{g}}{\sigma}\right) +
           \mu^+(\mu,\sigma,\ell,\alpha n)\,F\!\left(\frac{(t -\mu)\sqrt{g}}{\sigma}\right).
\end{eqnarray*}
}%
Proposition~\ref{pro:1} is proved.

\PPR{\ref{pro:2}}
Let $E_\xi$ be the event that the number of egoists' votes cast for a proposal is higher than~$\xi$. Let
\begin{gather}\label{e_Fxi}F_\xi = P(E_\xi).\end{gather}
Then~(see\:\cite{Che09PU})
$$
   F_\xi=\sum_{x=[\xi]+1}^{\ell}b(x\,|\,\ell) \approx F \left(-\frac{[\xi]+0{.}5-p\ell}{\sqrt{pq\ell}}\right)\!,
$$
where $p=F\left(\frac{\mu}{\sigma}\right),$ $q = F\left(-\frac{\mu}{\sigma}\right),$ and $b(x\,|\,\ell)
                                             = \binom{\ell}{x} p^x q^{\ell-x}$.

Using the notation $\text{\it MP}(A)=\mathrm{M}\big(\dG\,|\,A\big)P(A)$ we obtain
{ \hspace*{15mm} \arraycolsep=.3mm\begin{eqnarray}\label{e_probSum}
{\hspace*{10mm}\mathrm{M}}\big(\dG\big) &=&\text{\it MP}\big(G           \hspace*{-.35mm}\wedge\hspace*{-.35mm} E_{\gamma n}\big)
 \hspace*{-.35mm}\hspace*{-.35mm}+ \text{\it MP}\big(G           \hspace*{-.35mm}\wedge\hspace*{-.35mm}\overline E_{\gamma n}\big)
 \hspace*{-.35mm}+\hspace*{-.35mm} \text{\it MP}\big(\overline{G}\hspace*{-.35mm}\wedge \hspace*{-.35mm}E_{\alpha n}\big)
 \hspace*{-.35mm}+\hspace*{-.35mm} \text{\it MP}\big(\overline{G}\hspace*{-.35mm}\wedge\hspace*{-.35mm} \overline E_{\alpha n}\big) 
 \\\nonumber
 &=&\text{\it MP}\big(G          \wedge E_{\gamma n}\big)
 + \text{\it MP}\big(\overline{G}\wedge E_{\alpha n}\big),
\end{eqnarray}
}since the second and fourth terms of the original expression are zero (as sunder the corresponding conditions, the proposal is rejected and the participants' capital does not change). Introducing the notation
\begin{gather}\label{e_til}
\tilde t=\frac{(\mu - t)\sqrt{g}}{\sigma}
\end{gather}
and using \eqref{e_PQ}, \eqref{e_Fxi}, independence of the proposal components, the fact that $\dG$ is distributed as $N\big(\mu, \frac{\sigma^2}{g}\big),$ and formula (A.9) in \cite{Che06AiT} expressing $\mathrm{M}(\zeta\,|\,\zeta>t)$ for a normal $\zeta$, we obtain:
{\arraycolsep=.6mm\begin{eqnarray}\label{e_MP1}
 {\hspace*{10mm}\text{\it MP}}\big(G\hspace*{-.2mm}\wedge\hspace*{-.2mm} E_{\gamma n}\big)
 &\!=\!&
   \mathrm{M}\big(\dG\,|\,G\hspace*{-.2mm}\wedge\hspace*{-.2mm} E_{\gamma n}\big)\,P(G\hspace*{-.2mm}\wedge\hspace*{-.2mm} E_{\gamma n})
\hspace*{-.2mm} =\hspace*{-.2mm} \mathrm{M}\big(\dG\,|\,G\big)P(G)\,P(E_{\gamma n})\hspace*{-.35mm}
\hspace*{-15em}\\\nonumber
 &\!=\!&
   \mathrm{M}\big(\dG\,|\,\dG>t\big)F\!\left(\tilde t\xy\right) F_{\gamma n}
 = \left(\mu + \frac{\sigma f\!\left(\tilde t\xy\right)\!}{\sqrt{g}F\!\left(\tilde t\xy\right)}\right)F\!\left(\tilde t\xy\right)F_{\gamma n}
 \\\nonumber
 &\!=\!&
     \left(\mu P                           + \frac{\sigma f}{\sqrt{g}}\right)\!F_{\gamma n}.
    \end{eqnarray}
}

   Similarly,
    \begin{gather}
     \text{\it MP}\big(\overline{G}\wedge E_{\alpha n}\big)
     =\left(\mu Q                            - \frac{\sigma f}{\sqrt{g}}\right)\!F_{\alpha n}.
    \label{e_MP2}
    \end{gather}

    Substituting \eqref{e_MP1} and \eqref{e_MP2} into \eqref{e_probSum}, we obtain~\eqref{e_groupExp}.
    Proposition~\ref{pro:2} is proved.

\PTHO
    The expected capital increment of a randomly selected participant is equal to
    \begin{gather}\label{e_ramdomExp}
        \mathrm{M}(\tilde{d}) = \delta \mathrm{M}(\dE) + (1 - \delta)\mathrm{M}(\dG).
    \end{gather}

    Using Propositions $\ref{pro:1}$ and $\ref{pro:2}$ and the notation~\eqref{e_til} we now equate the derivative of the expression \eqref{e_ramdomExp} with respect to $t$ to zero:
   \begin{gather*}
    \delta\left[\mu^+(\mu,\sigma,\ell,\alpha n) - \mu^+(\mu,\sigma,\ell,\gamma n)\right]f(\tilde t\xy)\frac{\sqrt{g}}{\sigma}
     +(1-\delta)(F_{\gamma n} - F_{\alpha n})f(\tilde t\xy)\frac{\sqrt g}{\sigma}\left[-\mu + \frac{\sigma}{\sqrt g}\tilde t\right] = 0,
   \end{gather*} 
    which implies
    \begin{gather}\label{e_deri}
    \delta\,[\mu^+(\mu,\sigma,\ell,\alpha n) - \mu^+(\mu,\sigma,\ell,\gamma n)] - (1 - \delta)(F_{\gamma n} - F_{\alpha n})\,t = 0,
    \end{gather}
and consequently \eqref{e_generalOptimT} takes place.
Using \eqref{e_deri}, calculate the second derivative of \eqref{e_ramdomExp} at~$t_0$ denoting by $\tilde t_0$ the result of substituting $t_0$ into~\eqref{e_til}:
    \begin{gather*}
     f'(\tilde t_0)\frac{\sqrt{g}}{\sigma}\big(\delta\big[\mu^+(\mu,\sigma,\ell,\alpha n) - \mu^+(\mu,\sigma,\ell,\gamma n)] -
                                             (1 - \delta)(F_{\gamma n} - F_{\alpha n})\,t_0\big]\big) 
     +f(\tilde t_0) \frac{\sqrt{g}}{\sigma}(1 - \delta)(F_{\alpha n} - F_{\gamma n})\\ 
     =f(\tilde t_0) \frac{\sqrt{g}}{\sigma}(1 - \delta)(F_{\alpha n} - F_{\gamma n}).
    \end{gather*}
    This expression is negative. Consequently, $t_0$ is a point of maximum.
    Theorem 
    is proved.

\PCRO
If the  first condition is satisfied, then ${\gamma<0,}$ consequently, ${F_{\gamma n} = 1}$ and ${\mu^+(\mu,\sigma,\ell,\gamma n) = \mu}.$
If the second condition is      true, then ${F_{\alpha n} = \mu^+(\mu,\sigma,\ell,\alpha n) = 0}.$
Substituting these expressions into \eqref{e_generalOptimT} we complete the proof of Corollary~\ref{co:2}.


\revred{A.L. Fradkov}
\received{28.10.2016}
